\documentstyle[12pt]{article}
\begin{document}
\begin{titlepage}
{}~~~~~~~~~

\vspace{1.5cm}

\begin{center}

\Large{\underline{Gauge Invariances in the Proca Model}}\\
\vspace{1.5cm}
\large{A S Vytheeswaran}\\
{\em Department of Physics, Bangalore University, \\
   JnanaBharathi Campus, Bangalore 560 056~ INDIA}

\end{center}

\vspace{3cm}

\begin{abstract}

We show that the abelian Proca model, which is gauge non-invariant with 
second class constraints can be converted into gauge theories with first
class constraints. The method used, which we call Gauge Unfixing employs
a projection operator defined in the original phase space. This operator
can be constructed in more than one way, and so we get more than one
gauge theory. Two such gauge theories are the St\"uckelberg theory, and 
the theory of Maxwell field interacting with an antisymmetric tensor
field. We also show that the application of the projection operator 
does not affect the Lorentz invariance of this model.

\end{abstract}

\hfill

\begin{flushleft}
email:~~ctsguest@cts.iisc.ernet.in\\
{}~~~~~~~~~~libn@bnguni.kar.nic.in
\end{flushleft}
\end{titlepage}

{\noindent {\bf{1. \underline{Introduction}}}}

\vspace{3mm}

Hamiltonian systems with second class constraints$^{1}$ have been 
the subject matter of interest for sometime now. Although their 
existence has been known for long these constraints were regarded 
as merely serving to reduce the degrees of freedom, and hence are 
removed by using the Dirac bracket formalism. First class constraints 
on the other hand imply the presence of gauge invariance. 

Even though second class constraints by themselves do not imply 
gauge invariance in the corresponding systems, recent work$^{2,3}$ 
shows the possibility of underlying symmetries in such systems. 
These are revealed by converting the original second class system to
equivalent theories which have gauge invariance. In the language of 
constraints this means the new theories will now have first class 
constraints. 

Two methods are available for this conversion to equivalent gauge 
invariant theories. One is the BF method$^{2}$, which is basically 
formulated by extending the phase space of the original second class 
system. The other method is what we call Gauge Unfixing$^{3}$; this, 
unlike the BF method is formulated within the original phase space 
itself. The important step in this is the construction of a certain 
projection operator which defines the gauge theory. For a second 
class system this operator is not unique. It can be constructed 
in more than one way and so we can have more than one gauge theory, 
all equivalent to the original second class system.

The advantages of treating a second class constrained system in this 
manner are obvious. The new gauge theory can be studied using well 
established techniques like BRST, Dirac quantisation, etc,. Further 
under gauge fixing the new theory goes back to the old (gauge 
non-invariant) one for a specific gauge. But other gauges can also be 
used, gauges which might yield physically relevant theories. We know 
from the results of Faddeev and Fradkin-Vilkovisky$^{4}$ that these 
gauges are all equivalent. Apart from this freedom in choosing the 
gauge, we also have the freedom in choosing the appropriate projection 
operator and thus the appropriate gauge theory.

In this paper we consider the abelian Proca model in the light of  
the above method. This model has only second class constraints. The 
projection operators are constructed. For one choice of operator the 
resulting gauge theory has a trivial invariance, and the Hamiltonian 
is written entirely in terms of gauge invariant variables. The other 
choice for the projection operator gives a non-trivial gauge theory
which will be shown to lead to the (gauge invariant) St{\"{u}}ckelberg 
version$^{5}$ of the Proca model. Treated in a different manner, the 
Hamiltonian for this same non-trivial gauge theory leads to a model 
which has a massless antisymmetric tensor field interacting with the 
Maxwell field.

Many of these results have also been obtained by using the 
Batalin-Fradkin method$^{6,7}$ which, as we mentioned  earlier, is 
formulated in an extended phase space. However we emphasize that our 
results are obtained through Gauge Unfixing, which involves no extension 
of the phase space. In other words the gauge theories that we obtain 
can be found {\em within} the phase space of the original second 
class (Proca) theory.

We also look at the Poincar{\'{e}} invariance of the new gauge 
theories. The Proca model that we start with has a manifestly 
Lorentz invariant Lagrangian. In phase space its Poincar{\'{e}} 
generators obey the Poincar{\'{e}} algebra through Dirac brackets$^{1}$. 
We show that for either choice of the projection operator, these 
generators (even though they get modified by the projection 
operator) continue to obey the Poincar{\'{e}} algebra. The use 
of the projection operator thus does not affect Poincar{\'{e}} 
invariance.

In section 2 we introduce and summarize the gauge non-invariant Proca 
model. In section 3 we introduce the method of Gauge Unfixing and apply
it to the Proca model. The two choices of the first class constraint 
are dealt with separately as cases (i) and (ii). Section 4 is devoted 
to conclusions. In the appendix we give the properties of the projection
operator.

\vspace{5mm}

{\noindent{\bf{2.  {\underline {The Proca Model}}}}}
\vspace{3mm}

The abelian Proca model is a $ (3+1)$-dimensional theory given by the 
Lagrangian
$$
{\cal L} = -{\frac{1}{4}}F_{{\mu}{\nu}}F^{{\mu}{\nu}} + 
\frac{\displaystyle m^2}{\displaystyle 2}A_{\mu}A^{\mu},
\eqno(2.1)
$$
with $\, m $ the mass of the $ A_{\mu} $ field. As usual 
$F_{{\mu}{\nu}} = {\partial_{\mu}}A_{\nu} - {\partial_{\nu}}A_{\mu},
$ and $ g_{\mu\nu} = {\em diag}(+,-,-,-). $

In phase space, we have the momenta $ \pi_{\mu}(x) $  conjugate to
the $ A^{\mu}(x) $ and the canonical Hamiltonian (after ignoring a
total derivative term which arises in the Legendre transformation)
$$
H_c = {\displaystyle \int d^3x ~{\cal H}_c} =  {\displaystyle \int 
d^3x \left ( \frac{1}{2}{\pi_i}{\pi_i} 
+ {\frac{1}{4}}F_{ij}F_{ij} - \frac{m^2}{2}(A_0^2 - A_i^2) + 
A_0({\partial_i}\pi_i)\right ),}
\eqno(2.2)
$$
with $ \pi_i = -F_{0\,i} $. There are two second class constraints
$$
\begin{array}{rcl}
Q_1 & = & \pi_0(x) \approx 0,\\
Q_2 & = & (-{\partial_i}\pi_i + {m^2}A_0)(x) \approx 0,
\end{array}
\eqno(2.3)
$$
where $ Q_1 $ is the primary constraint and $ Q_2 $ the secondary 
constraint. These two constraints together define the surface $ 
\sum_2 $ in the phase space. Their second class nature is seen by 
their non-zero Poisson brackets
$$
\{ Q_1(x), Q_2(y)\} = -{m^2}\,\delta(x-y).           
\eqno(2.4)
$$
We thus have a $ 2 \times 2 $ matrix $ E $ with elements $ E_{ab} 
= \{ Q_a(x), Q_b(y) \}  ~~(a,b = 1,2), $
$$
\left( \begin{array}{c}
       0\;\;\;{-m^2}\\
	 \!\!{m^2}\:\;\;\;0
      \end{array}\right )\;,
\eqno(2.5)
$$
which has a non-zero determinant and hence an inverse $ E^{-1} $
everywhere in the phase space. The constraints (2.3) can be eliminated 
by replacing Poisson brackets (PBs) by Dirac brackets (DBs). For any 
two phase space functions $ B $ and $ C $,
$$
\begin{array}{rcl}
\{ B(x), C(y) \}_{DB} & = & \{ B(x), C(y) \}_{PB}\\
&   & - {\displaystyle \int d^3u\, d^3v\, \{ 
B(x), Q_a(u) \}_{PB}E^{-1}_{ab}(u,v)\{Q_b(v), C(y)\}_{PB}}.
\end{array}
\eqno(2.6)
$$
By construction the Dirac bracket of any variable with either of 
the constraints (2.3) is exactly zero. Further we have
$$
\begin{array}{rcl}
\{ A^i(x), \,\pi_j(y) \}_{DB} & = & \delta^i_{j}\delta(x-y),\\
\{A^i(x), \,A^j(y)\}_{DB} & = & \{\pi_i, \pi_j \}_{DB} \,= 0,\\
\{A_0(x),\, \pi_i(y)\}_{DB} & = & 0,\\
\{A_0(x), A_i(y)\}_{DB} & = & \frac{\displaystyle 1}{\displaystyle
m^2}\partial_{ix}\delta(x-y).
\end{array}
\eqno(2.7)
$$
Thus the $ A^i $ and the $ \pi_j $ continue to remain canonical 
conjugate pairs. However from the last equation in (2.7), we see that 
$ A_0 $ is no longer independent of the $ \pi_i $. This equation is 
consistent with taking $ Q_2 = 0 $ as a strong equation$^{1}$ and 
replacing $ A_0 $ by $ \frac{\displaystyle {\vec \nabla}\cdot{\vec 
\pi}}{\displaystyle m^2}. $ Using $ Q_2 = 0, $ the canonical 
Hamiltonian (2.2) becomes 
$$
H_c = {\displaystyle \int d^3x \left ({\frac{1}{2}}{\pi_i}{\pi_i} + 
\frac{m^2}{2}{A_i}^2 + \frac{\displaystyle F_{ij}F_{ij}}{
\displaystyle 4} + \frac{\displaystyle (\partial_i\pi_i)^2}{
\displaystyle 2m^2}\right )}.
\eqno(2.8)
$$

The Lagrangian $ L $ in (2.1) is manifestly Lorentz invariant. To
verify Poincar{\'{e}} invariance of the model in phase space, we 
look at the components of the energy-momentum and angular momentum
tensors
$$
\begin{array}{rcl}
{\cal T}_{0\mu} & = & -F_{0\alpha}(\partial_{\mu}A^{\alpha}) - 
g_{0\mu}{\cal L},\\
{\cal M}_{0\mu\nu} & = & x_{\mu}{\cal T}_{0\nu} - x_{\nu}{\cal 
T}_{0\mu} + \pi_{\mu}A_{\nu} - \pi_{\nu}A_{\mu}.
\end{array}
$$
The Poincar{\'{e}} group generators $ P_{\mu} = \int d^3x~{\cal 
T}_{0\mu} $ and $ M_{\mu\nu} = \int d^3x ~{\cal M}_{0\mu\nu} $
in phase space are then
$$
\begin{array}{rcl}
P_0 & = & {\displaystyle \int d^3x ~\left (\frac{1}{2}\pi_i\pi_i + 
\frac{1}{4}F_{ij}F_{ij}
- \frac{m^2}{2}[A_0^2 - A_i^2] + A_0(\partial_i\pi_i)\right ),}\\
P_i & = & {\displaystyle \int d^3x ~\left ({\vec \pi}\cdot({
\partial_i}{\vec A})\right )},\\
M_{0i} & = & {\displaystyle \int d^3x ~\left (x_0{\vec \pi}\cdot({
\partial_i}{\vec A})- x_i{\cal H}_c + {\pi_0}A_i \right )},\\
M_{ij} & = & {\displaystyle \int d^3x ~\left (x_i\,{\vec \pi}\cdot({
\partial_j}{\vec A}) - x_j\,{\vec \pi}\cdot({\partial_i}{\vec A})
+ {\pi_i}A_j - {\pi_j}A_i \right )},
\end{array}
\eqno(2.9)
$$
where in the first line, a total derivative term is ignored. 
Furthermore, though the term $ - \pi_iA_0 $ is present in $ {\cal 
M}_{00i}, $ it is absent in $ M_{0i}. $ This is so because the 
substitution  (and rewriting) of the expression for $ {\cal T}_{00} 
$ in $ M_{0i} $ gives rise to a term $ +\pi_iA_0 $ (apart from a 
total derivative) which cancels the $ -\pi_iA_0 $ already present.
The resulting expression for $ M_{0i} $ is the one shown in (2.9), 
with $ {\cal H}_c $ is the Hamiltonian density given in (2.2).
Using the Dirac brackets (2.6) we find on the surface $ \sum_2, $
$$
\begin{array}{rcl}
\{P_{\mu}, P_{\nu} \}_{DB} & = & 0,\\
\{M_{{\mu}{\nu}}, P_{\lambda} \}_{DB} & = & - g_{{\mu}{\lambda}}P_{\nu} 
+ g_{{\nu}{\lambda}}P_{\mu},\\
\{M_{{\mu}{\nu}}, M_{{\sigma}{\rho}} \}_{DB} & = & - g_{{\mu}{\sigma}}
M_{{\nu}{\rho}} + g_{{\nu}{\sigma}}M_{{\mu}{\rho}} + g_{{\mu}{\rho}}
M_{{\nu}{\sigma}} - g_{{\nu}{\rho}}M_{{\mu}{\sigma}}.
\end{array}
\eqno(2.10)
$$
It is important to note (for later purposes) that the right hand sides
of (2.10) (apart from total derivatives) also have terms involving the
constraints (2.3), which have been put to zero here.
The Poincar{\'{e}} algebra (2.10) thus confirms the Poincar{\'{e}} 
invariance of the Proca model in the Hamiltonian formulation.

\vspace{5mm}

{\noindent {\bf{3.  {\underline{Gauge Unfixing}}}}}

\vspace{3mm}

We now derive the underlying symmetries of the Proca model using the 
gauge unfixing method$^{3}$. For this we first note from (2.5) that 
each of the constraints in (2.3) is first class (i.e., has zero PB) 
with itself, but they are second class with respect to each other.
Thus each is like a gauge fixing constraint to the other. Now if 
either of these constraints is retained and the other no 
longer considered a constraint, then we have a system with only a 
first class constraint. Accordingly we have two choices for our 
first class constraint. We consider these one by one.

\vspace{5mm}

{\large {\noindent \bf{{{Case}} (i)}}}

\vspace{3mm}

We redefine the constraints (2.3) as
$$
\begin{array}{rcl}
\chi(x) & = & -\frac{\displaystyle 1}{\displaystyle m^2}Q_1(x), \\
\psi(x) & = & ~~~~~~~Q_2(x),
\end{array}
\eqno(3.1)
$$
so that, from (2.4) $ \chi $ and $ \psi $ form a canonical conjugate 
pair. We now choose $ \chi \cong 0 $ as our first class constraint,
and no longer consider $ \psi \approx 0. $ The dynamics will now be 
relevant on a new constrained surface $ \sum_1 $ defined by only $ 
\chi \cong 0 $ (the equality sign is changed from $ \approx $ to $ 
\cong $). In order that we have a gauge theory with transformations 
generated by $ \chi, $ relevant physical quantities must be gauge 
invariant. In particular the Hamiltonian $ H_c $ of (2.2) is not gauge
invariant, $ \{ \chi, H_c \} \not\cong 0 $ (on $ \sum_1 $). Hence to 
get gauge invariant observables, we define a projection operator
$$
I\!\!P =~ : e^{-{\displaystyle \int{d^3x} ~{\psi\,}{\hat \chi}}} :\,\,,
\eqno(3.2)
$$where for any phase space functional $ B, $ we have $ {\hat \chi}B 
\equiv \{ \chi, B \}. $ In applying (3.2) we adopt a particular 
ordering$^{3}$; when $ I\!\!P $ acts on any $ B, \psi $ should 
always be outside the Poisson bracket. We thus have the gauge 
invariant quantity $ {\widetilde B}(x) $
$$
\begin{array}{rcl}
{\widetilde B}(x) & = & :e^{-{\displaystyle \int d^3x~ \psi\,{\hat 
\chi}}}:B(x)\\
		  & = & B(x) - {\displaystyle \int d^3y~ \psi(y) \{ 
\chi(y), B(x)\}}\\
&   & + 
\frac{\displaystyle 1}{\displaystyle 2!}{\displaystyle \int d^3y d^3z~ 
\psi(y)\psi(z) \{ \chi(y), \{ \chi(z), B(x) \} \} } - .............+....
\end{array}
\eqno(3.3)
$$
In particular, the gauge invariant Hamiltonian will be, using (2.2) and 
(3.3)
$$
\begin{array}{rcl}
{\widetilde H}_c & = & H_c - {\displaystyle \int d^3x~ \psi(x)
(-\frac{\displaystyle 1}{\displaystyle m^2})\psi(x)}
+ \frac{1}{2\!}{\displaystyle \int d^3xd^3y~ \psi(x)\psi(y) 
(\frac{\displaystyle -1}{\,\displaystyle m^2})\delta(x-y)}\\
		 & = &\rule{0mm}{9mm} {\displaystyle \int d^3x \left 
(\frac{\displaystyle {\vec \pi}^2}{\displaystyle 2} + 
A_0{\vec \nabla}\cdot{\vec \pi} + \frac{1}{4}F_{ij}F_{ij} - 
\frac{m^2}{2}
(A_0^2 - {\vec A}^2) + \frac{\displaystyle \psi^2}{\displaystyle 
2m^2}\right )}.
\end{array}
\eqno(3.4)
$$
It can be checked that $ \{ \chi(x), {\widetilde H}_c \} = 0 $. Thus 
$ \chi \cong 0 $ and $ {\widetilde H}_c $ describe a consistent gauge 
theory. This gauge theory goes back to the original Proca model when 
we consider and use $ \psi \approx 0 $ as the gauge fixing condition.

The $ \chi $ is the generator of gauge transformations. It can be checked
that the $ A^i $ and the $ \pi_i ~(i = 1,2,3) $ are all gauge invariant. 
However $ A_0 $ is not, since $ A_0 \rightarrow A_0^{\prime} = A_0 + 
\lambda, $ for infinitesimal gauge transformations. Here $ \lambda $ 
is the transformation parameter.

The Hamiltonian $ {\widetilde H}_c $ though gauge invariant, involves 
gauge non-invariant fields. Using the explicit form $ \psi = \left 
(-{\vec \nabla}\cdot{\vec \pi} + m^2\,A_0 \right ), $ it can be 
rewritten in terms of only gauge invariant fields,
$$
{\widetilde H}_c =  {\displaystyle \int d^3x \left (\frac{\displaystyle
\pi_i^2}{\displaystyle 2} + \frac{\displaystyle m^2\,A_i^2}{\displaystyle 
2} + \frac{\displaystyle F_{ij}F_{ij}}{\displaystyle 4} + 
\frac{\displaystyle (\partial_i \pi_i)^2}{\displaystyle 2m^2} \right )},
\eqno(3.5)
$$
where the fields $ A^i(x) $ and $ \pi_j(x) $ continue to form 
canonical conjugate pairs. Note that $ {\widetilde H}_c $ in (3.5) is 
just the Dirac bracket Hamiltonian (2.8) of the original Proca theory. 

We now look at the Poincar{\'{e}} invariance of the new gauge 
theory. In order that the Poincar{\'{e}} group generators be 
physical observables, they must be gauge invariant with respect to $ 
\chi. $ To obtain these, we first apply $ I\!\!P $ on the quantities
$ P_{\mu},\,M_{\mu\nu} $ of (2.9),
$$
\begin{array}{rcl}
I\!\!P(P_0) & = & {\widetilde {P_0}} ~~= P_0 + {\displaystyle 
\int d^3x \left (\frac{\displaystyle \psi^2}{\displaystyle 
2m^2}\right )},\\
I\!\!P(P_i) & = & {\widetilde {P_i}} ~~= P_i,\\
I\!\!P(M_{0i}) & = & {\widetilde M}_{0i} = M_{0i} - {\displaystyle
\int d^3x \left (\frac{\displaystyle x_i\psi^2}{\displaystyle 
2m^2}\right )}\\
I\!\!P(M_{ij}) & = & {\widetilde M}_{ij} = M_{ij}.
\end{array}
\eqno(3.6)
$$
These projected quantities can be verified to be gauge invariant.
Thus in order that they be gauge invariant, $ P_0 $ and $ M_{0i} $
get modified. The Poincare algebra is verified by looking at the 
Poisson brackets of the projected quantities (3.6). To this end we 
use certain properties of the projection operator (see appendix). 
Using (A.6), the Dirac brackets (2.10), (A.5) and (A.4), we find 
on the surface $ \sum_1 (\,\chi \cong 0) $
$$
\begin{array}{rcl}
\{ {\widetilde {P_{\mu}}}, {\widetilde {P_{\nu}}} \} & \cong & 0,\\
\{ {\widetilde M}_{\mu\nu}, {\widetilde {P_{\lambda}}} \} & \cong & 
-g_{\mu\lambda}{\widetilde {P_{\nu}}} + g_{\nu\lambda}{\widetilde 
{P_{\mu}}},\\
\{ {\widetilde M}_{\mu\nu}, {\widetilde M}_{\sigma\rho} \} & 
\cong & -g_{\mu\sigma}{\widetilde M}_{\nu\rho} + g_{\nu\sigma}
{\widetilde M}_{\mu\rho} + g_{\mu\rho}{\widetilde M}_{\nu\sigma}
- g_{\nu\rho}{\widetilde M}_{\mu\sigma},
\end{array}
\eqno(3.7)
$$
which shows that the Poincar{\'{e}} algebra is not affected by the 
projection
operator $ I\!\!P $ (3.2). In this context it must be noted that it is 
necessary here to have $ I\!\!P-$projected Poincar{\'{e}} 
generators instead 
of the old ones (2.9). If we consider the old generators (2.9), then 
their PB or DB algebra (2.10) will in general involve both $ \chi $ 
and $ \psi $, which can both be put to zero (surface $\sum_2$) {\em 
only in} the original second class theory. In our new gauge theory, 
only $ \chi $ can be put to zero (on $ \sum_1 $), and so the old 
generators (2.9) no longer give the Poincar{\'{e}} algebra. But 
if instead the 
$ I\!\!P-$projected quantities (3.6) are used, even if their Poisson 
brackets give extra terms involving $ \psi, $ these get eliminated 
due to the property (A.4), and Poincar{\'{e}} algebra is obtained.

The inverse Legendre transformation for the Hamiltonian $ {\widetilde 
H}_c $ (3.4) will result in a Lagrangian which is not manifestly Lorentz
invariant. We do not consider this here.

\vspace{7mm}

{\large \bf{{\noindent{{{Case}} (ii)}}}}

\vspace{3mm}

To consider a different choice of first class constraint, we reclassify
the constraints (2.3) as
$$
\begin{array}{rcl}
\chi^{\,\prime}(x) & = &  \frac{\displaystyle 1}{\displaystyle 
m^2}Q_2(x) = \frac{\displaystyle 1}{\displaystyle m^2}\left (-{\vec
\nabla\,}\cdot{\!\vec \pi} + m^2A_0\right )\\
\psi^{\prime}(x) & = &\rule{0mm}{7mm} Q_1(x) = \pi_0(x),
\end{array}
\eqno(3.8)
$$
which, as in the earlier classification (3.1), form canonical conjugate
pairs. We choose $ \chi^{\,\prime} \cong^{\prime} 0 $ (note the change
in equality sign) to be our first class constraint, and disregard $ 
\psi^{\,\prime} \approx 0. $ Then $ \chi^{\,\prime} \cong^{\prime} 0 
$ will define a new constrained surface $ \sum_1^{\prime}, $ different 
{}from the earlier $ \sum_2 $ and $ \sum_1. $ Our new gauge theory is 
now to be defined on this new $ \sum_1^{\,\prime}. $

As in case (i), we must have observables gauge invariant under gauge 
transformations generated here by $ \chi^{\,\prime}. $ Quantities
like the second class Hamiltonian $ H_c $ of (2.2) do not in general 
satisfy this requirement. Further the Hamiltonian $ 
{\widetilde H}_c $ of (3.4), which was gauge invariant in the earlier
case (i) is not so here, $ \{ \chi^{\,\prime}, {\widetilde H}_c \} 
\,{\not\cong}^{\prime} \,0. $ Hence we define and construct a new 
projection operator
$$
\begin{array}{rcl}
I\!\!P^{\,\prime} & = & : e^{-{\displaystyle \int d^3x ~\psi^{\prime}(x)
{\hat \chi}^{\prime}(x)}} :\\
I\!\!P^{\,\prime}(B) & \equiv & {\widetilde B}^{{}^{^{\prime}}}\\
{\hat \chi}^{\,\prime}B & = & \{ \chi^{\,\prime}, B \},
\end{array}
\eqno(3.9)$$
where $ B $ is any phase space functional. Again, as in (3.2), we have 
here a particular ordering --- the $ \psi^{\prime} $ is 
always outside the PBs occuring in the series expansion of $ 
I\!\!P^{\,\prime}(B). $ 

It must be noted that the $ I\!\!P^{\,\prime} $ and $ {\hat \chi}^{\,\prime} 
$ in (3.9) are {\em not\,} the same as the $ I\!\!P $ and $ {\hat 
\chi} $ in (3.2). Thus the gauge theory defined by $ I\!\!P^{\,\prime} 
$ is in general different from the one defined by $ I\!\!P. $ For 
instance, the gauge invariant Hamiltonian $ I\!\!P^{\,\prime}(H_c) 
(={\widetilde H_c}^{{}^{^{\,\prime}}}) $ here is different from 
the $ {\widetilde H}_c $ of eqn.(3.4),
$$
\begin{array}{rcl}
{{\widetilde H}_c}^{{}^{^{\,\prime}}}
& = & H_c - \displaystyle \int d^3x ~\left [\psi^{\,\prime}({\vec 
\nabla}\cdot{\vec A}) + \frac{\displaystyle 1}{\displaystyle 
2m^2}\psi^{\prime}{\vec \nabla}^2\psi^{\prime}\right ]\\
& = &\rule{0mm}{9mm} {\displaystyle \int d^3x \,\left 
[\frac{\displaystyle \pi_i^2}{\displaystyle 2} + A_0({\partial_i\pi_i}) 
+ \frac{\displaystyle 1}{\displaystyle 4}F_{ij}F_{ij} - 
\frac{\displaystyle m^2}{\displaystyle 2}(A_0^2 - A_i^2)\right. }\\
&   &~~~~~~~~~~~~{\left. - \pi_0({\vec \nabla\,}\cdot{\!\vec A})
+ \frac{1}{2m^2}({\vec \nabla}\pi_0)^2 \right ]}.
\end{array}
\eqno(3.10)
$$
It can be verified that $ \{ \chi^{\,\prime}(x), {\widetilde 
{H}_c}^{{}^{^{\,\prime}}}
\} = 0. $ Thus $ \chi^{\,\prime} \cong^{\prime} 0 $ and $ {{\widetilde
H}_c}^{{}^{^{\,\prime}}} $ define our new gauge theory. This goes 
back to the Proca theory under the gauge condition $ 
\psi^{\,\prime} \approx 0. $ The Hamiltonian $ {\widetilde 
{H}_c}^{{}^{^{\,\prime}}} $ goes back to the second class Hamiltonian 
(2.2).

The gauge transformations are generated by $ \chi^{\,\prime}, $ and 
unlike in case (i) (where $ A^i $ and $ \pi_j $ were gauge invariant), 
here the gauge invariant fields are $ A_0 $ and $ \pi_i $. As for the 
remaining fields, we have, for a transformation parameter $ \mu(x) $
$$
\begin{array}{rcl}
A_i \rightarrow A_i^{\prime} & = & A_i + {\displaystyle \int d^3x ~
\mu(x)\{ A_i, \chi^{\,\prime}(x) \}}\\
& = & A_i - \frac{\displaystyle 1}{\displaystyle m^2}(\partial_i{\mu}),\\
\pi_0 \rightarrow \pi_0^{\,\prime} & = & \pi_0 - \mu.
\end{array}
\eqno(3.11)
$$
It can also be verified explicitly using (3.11) that $ {\widetilde
{H}_c}^{{}^{^{\,\prime}}} $ is gauge invariant.

\vspace{4mm}

Before we look further at the Hamiltonian $ {{\widetilde 
H}_c}^{{}^{^{\,\prime}}}$, we look for Poincar{\'{e}} invariance 
in this new gauge theory. As
in case (i), the group generators must be gauge invariant, this
time with respect to $ \chi^{\,\prime}. $ It can be seen that neither 
the quantities (2.9) of the second class Proca theory, nor the quantities
(3.6) have zero PBs with $ \chi^{\,\prime}. $ Hence we apply the operator
$ I\!\!P^{\,\prime} $ (3.9) on all the quantities $ P_{\mu}, 
M_{\mu\nu} $ of (2.9). Noting from (3.9) that the operation of $ 
I\!\!P^{\prime} $ results in a series, we get the gauge invariant
quantities
$$
\begin{array}{rcl}
{\widetilde {P_0}}^{\prime} & = & {\displaystyle \int d^3x 
\left (\frac{\displaystyle{\pi_i^2}}{\displaystyle 2} + 
A_0(\partial_i\pi_i) - \frac{\displaystyle m^2}{
\displaystyle 2}(A_0^2 - {A_i}^2)
+ \frac{1}{4}F_{ij}^2 \right. }\\
&  & {\left. ~~~~~~~- \pi_0({\vec \nabla}\cdot{\vec A}) + 
\frac{\displaystyle 1}{\displaystyle 2m^2}({\vec \nabla}\pi_0)^2 \right 
)},\\
{\widetilde {P_i}}^{\prime} & = & \rule{0mm}{8mm} P_i - {\displaystyle 
\int d^3x ~\frac{\displaystyle 1}{\displaystyle 
m^2}\psi^{\,\prime}[-\partial_i({\vec \nabla}\cdot{\vec \pi})]}\\
& = & {\displaystyle \int d^3x \left (\pi_{\mu}\partial_i{A^{\mu}} 
- \psi^{\prime}\partial_i{\chi^{\,\prime}}\,\right )},\\
{{\widetilde {M}_{0i}}}^{{}^{^{\,\prime}}} & = & \rule{0mm}{8mm} M_{0i} 
+ {\displaystyle \int d^3x \left (\frac{\displaystyle x_0}
{\displaystyle m^2}\pi_0\partial_i({\vec \nabla}\cdot{\vec \pi}) 
+ x_i\pi_0({\vec \nabla}\cdot{\vec A}) + \frac{\displaystyle 
x_i}{\displaystyle 2m^2}\pi_0{\vec \nabla}^2\pi_0 \right )}\\
& = & \rule{0mm}{7mm}{\displaystyle \int d^3x \left 
[x_0\pi_{\mu}\partial_i{A^{\mu}} - 
x_0\psi^{\prime}\partial_i{\chi^{\,\prime}} - x_i{\widetilde 
{\cal H}_c}^{{}^{^{\,\,\,\prime}}} + \pi_0A_i\right ]},\\
{{\widetilde {M}_{ij}}}^{{}^{^{\,\prime}}} & = & \rule{0mm}{8mm}
M_{ij}+ {\displaystyle \int d^3x~ \psi^{\,\prime}(x_i\partial_j 
- x_j\partial_i)(\frac{\displaystyle{\vec \nabla}\cdot{\vec 
\pi}}{\displaystyle m^2})}\\
& = &\rule{0mm}{7mm}{\displaystyle \int d^3x \left 
(x_i\,\pi_{\mu}\partial_jA^{\mu} - x_j\,\pi_{\mu}\partial_iA^{\mu} 
+ \pi_i A_j - \pi_j A_i\right.}\\
&   & {\left.~~~~~~~~~~~~~~~-\psi^{\,\prime}(x_i\partial_j - 
x_j\partial_i)\chi^{\,\prime}\right )},
\end{array}
\eqno(3.12)
$$
where we have used (2.9), (2.3), (3.8) and the Hamiltonian density
$ {\widetilde {\cal H}_c}^{{}^{^{\,\prime}}} $ of (3.10). Also we have 
ignored total derivative terms in writing the expressions for 
$ {\widetilde P}_0^{{}^{^{\,\prime}}} $ and $ {\widetilde 
M}_{0i}^{{}^{^{\,\prime}}}. $ 
It can be verified that the above projected quantities
are indeed gauge invariant with respect to $ \chi^{\,\prime}. $

We now verify the Poincar{\'{e}} algebra. The old generators $ 
P_{\mu},M_{\mu\nu} $ of (2.9) will not serve this purpose here. 
This is because, as mentioned in section 2, the Dirac brackets 
(2.10) will in general involve extra terms involving both the 
$ \chi^{\,\prime} $ and $ \psi^{\,\prime} $. In the present 
gauge theory, only the $ \chi^{\,\prime} $ can be put to zero 
(surface $ \sum_1^{\,\prime} $) and {\em not} the $ \psi^{\,\prime}, 
$ in which case we will not have the Poincar{\'{e}} algebra.

For similar reasons the $ {\widetilde P}_{\mu} $ and $ {\widetilde 
M}_{\mu\nu} $ of (3.6) which obeyed the Poincar{\'{e}} algebra
in the gauge theory of case(i), cannot do so here. The PBs (3.7) 
among $ {\widetilde P}_{\mu} $ and $ {\widetilde M}_{\mu\nu} $ 
involved extra terms in $ \pi_0 ( = \psi^{\,\prime}), $ which 
cannot be put to zero here. Consequently we are left with 
verifying if the $ {\widetilde {P}_{\mu}}^{{}^{^{\,\prime}}}, 
{}~{\widetilde {M}_{\mu\nu}}^{{}^{^{\,\prime}}} $  of (3.12) 
satisfy the Poincar{\'{e}} algebra.

As in case (i), we use the properties of the projection operator 
given in the appendix. We use the Dirac brackets (2.10), and using 
(A.6), (A.5) and (A.4) we eliminate the extra terms in $ \psi^{\prime}. 
$ We thus get on the constraint surface $ \sum_1^{\,\prime} $
$$
\begin{array}{rcl}
\{ {\widetilde {P}_{\mu}}^{{}^{^{\,\prime}}}, {\widetilde 
{P}_{\nu}}^{{}^{^{\,\prime}}} \}
& = & I\!\!P^{\,\prime}\left (\{ P_{\mu}, P_{\nu} \}_{DB} \right ) 
~\cong^{\,\prime} 0,\\
\{ {\widetilde {M}_{\mu\nu}}^{{}^{^{\,\prime}}}, {\widetilde 
{P}_{\lambda}}^{{}^{^{\,\prime}}} \} & = & I\!\!P^{\,\prime}\left 
(\{ M_{\mu\nu}, P_{\lambda} \}_{DB}\right ) ~~\cong^{\,\prime} - 
g_{\mu\lambda}{\widetilde {P}_{\nu}}^{{}^{^{\,\prime}}} + 
g_{\nu\lambda}{\widetilde {P}_{\mu}}^{{}^{^{\,\prime}}},\\
\{ {\widetilde {M}_{\mu\nu}}^{{}^{^{\,\prime}}}, {\widetilde 
{M}_{\sigma\rho}}^{{}^{^{\,\prime}}} \} & = & I\!\!P^{\,\prime}\left 
(\{ M_{\mu\nu}, M_{\sigma\rho} \}_{DB}\right )\\
& = & -g_{\mu\sigma}{\widetilde {M}_{\nu\rho}}^{{}^{^{\,\prime}}} + 
g_{\nu\sigma}{\widetilde {M}_{\mu\rho}}^{{}^{^{\,\prime}}} + 
g_{\mu\rho}{\widetilde {M}_{\nu\sigma}}^{{}^{^{\,\prime}}} - 
g_{\nu\rho}{\widetilde {M}_{\mu\sigma}}^{{}^{^{\,\prime}}}.
\end{array}
\eqno(3.13)
$$
Thus the Poincar{\'{e}} algebra is satisfied and $ {\widetilde 
{P}_{\mu}}^{{}^{^{\,\prime}}},\, {\widetilde 
{M}_{\mu\nu}}^{{}^{^{\,\prime}}} $ are the generators of this group in 
the gauge theory defined by $ \chi^{\,\prime} \cong^{\,\prime} 0. $ 
The application of $ I\!\!P^{\,\prime} $ thus does not affect the 
Poincar{\'{e}} invariance of the Proca model.

\vspace{5mm}

We now return to the gauge invariant Hamiltonian $ {{\widetilde 
H}_{c}}^{{}^{^{\,\prime}}} $ of (3.10). The equations of motion are
$$
\begin{array}{rcl}
{\dot A}_0 & = & -{\,\vec \nabla}\cdot{\!\vec A} - 
\frac{\displaystyle {{\vec \nabla}^2\psi^{\,\prime}}}{\displaystyle 
m^2},\\
{\dot \pi}_0 & = &\rule{0mm}{8mm} Q_2 = -m^2\chi^{\,\prime} 
~~~\cong^{\prime} 0,\\
{\dot A}_i & = &  -\pi_i + \partial_i{A_0},\\
{\dot \pi}_i & = & \partial_j{F_{ji}} + m^2A_i - 
\partial_i{\psi^{\,\prime}}.
\end{array}
\eqno(3.14)
$$
We once again see the equivalence of this gauge theory (case(ii)) with 
the original Proca model. Under the gauge fixing condition $ 
\psi^{\,\prime} \approx 0, $ the equations (3.14) go back to the 
equations of motion for the Proca model.

We now consider the passage from $ {\widetilde {H}_c}^{{}^{^{\,\prime}}} 
$ to the Lagrangian formulation. Using $ {\dot \pi}_0 = Q_2 $ from 
(3.14), we rewrite $ {\widetilde {H}_c}^{{}^{^{\,\prime}}} $ as
$$
\begin{array}{rcl}
{\widetilde {H}_c}^{{}^{^{\,\prime}}} & = & {\displaystyle \int d^3x 
\left (\frac{\displaystyle {\vec \pi}^2}{\displaystyle 2} + \left 
[A_0 - \frac{\displaystyle 1}{\displaystyle m^2}\partial_0{\pi_0}\right 
]({\vec \nabla}\cdot{\vec \pi}) + \frac{1}{4}F_{ij}F_{ij} - 
\pi_0{{\,\vec \nabla}\cdot{\!\vec A}}\right. }\\
&   & ~~~~~~~~~\left. - \frac{\displaystyle m^2}{\displaystyle 
2}\left [(A_0 - \frac{1}{m^2}\partial_0{\pi_0})^2 - {\vec A}^2 
\right ] - \frac{\displaystyle 1}{\displaystyle 2m^2}\left 
[(\partial_{0}\pi_0)^2 - ({\vec \nabla}\pi_0)^2\right ]\right )\\
& = &\rule{0mm}{10mm} {\displaystyle \int d^3x \left 
(\frac{\displaystyle {\vec \pi}^2}{\displaystyle 2} + A_0^{\prime}{\vec 
\nabla}\cdot{\vec \pi} + \frac{1}{4}F_{ij}F_{ij} - 
\frac{m^2}{2}({A_0^{\prime}}^2 - {\vec A}^2)\right. }\\
&   & ~~~~~~~~~~~{\left. - \pi_0{\vec \nabla}\cdot{\vec A} - 
\frac{\displaystyle 1}{\displaystyle 
2m^2}(\partial_{\mu}{\pi_0})^2)\right )},
\end{array}
\eqno(3.15)
$$
where we have called $ (A_0 - \frac{\displaystyle 1}{\displaystyle
m^2}\partial_0{\pi_0}) = A_0^{\prime}. $ The equation of motion for 
$ A_0^{\prime} $ is the same as for $ A_0, $ because of $ {\dot 
\pi}_0 = m^2\chi^{\,\prime}(\cong^{\,\prime} 0). $ Further in eqn.(3.14) 
for $ {\dot A}_i, ~\partial_i{A_0} $ becomes $ \partial_i{A_0^{\prime}}. 
$ Using (3.15) and ignoring the prime on $ A^{\prime}_0, $ we write 
the Lagrangian
$ L = {\displaystyle \int d^3 x \left ( \pi_0{\dot A}_0 
+ \pi_i{\dot A}^i - {{\widetilde {\cal H}}_c}^{{}^{^{\,\,\prime}}}\right 
)} $ as 
$$
\begin{array}{rcl}
L & = &\rule{0mm}{8mm} {\displaystyle \int d^3x \left (\pi_0
\partial_{\mu}A^{\mu} - \pi_i( -\pi_i + \partial_i{A_0}) - 
\frac{\displaystyle \pi_i^2}{\displaystyle 2} 
- \frac{1}{4}F_{ij}F_{ij}\right. }\\
&   &~~~~~~~~~~~{\left. - A_0(\partial_i\pi_i) + 
\frac{\displaystyle m^2}{\displaystyle 2}A_{\mu}A^{\mu} + 
\frac{\displaystyle 1}{\displaystyle 2m^2}(\partial_{\mu}\pi_0)^2\right 
)}\\& = &\rule{0mm}{9mm} {\displaystyle \int d^3x \left 
(\pi_0\partial_{\mu}A^{\mu} - \frac{1}{4}F_{\mu\nu}F^{\mu\nu} + 
\frac{\displaystyle m^2}{\displaystyle 2}A_{\mu}A^{\mu} + 
\frac{\displaystyle 1}{\displaystyle 2m^2}(\partial_{\mu}\pi_0)^2\right 
)}.
\end{array}
\eqno(3.16)
$$
If we now consider the $ \pi_0 $ (or $ \psi^{\prime} $) to be a new
field appearing in the Lagrangian, and then rescale $ \pi_0 $ to
$ \theta = \frac{\displaystyle \!\!-1}{\displaystyle m^2}\pi_0, $ we 
can rewrite $ L $ as
$$
\begin{array}{rcl}
L & = & {\displaystyle \int d^3x \left (- \frac{1}{4}F_{\mu\nu}F^{\mu\nu} 
+ \frac{m^2}{2}A_{\mu}A^{\mu} - m^2\theta\partial_{\mu}A^{\mu} + 
\frac{\displaystyle m^2}{\displaystyle 2}(\partial_{\mu}\theta)^2\right 
)}\\
& = &\rule{0mm}{8mm} {\displaystyle \int d^3x \left ( - 
\frac{1}{4}F_{\mu\nu}F^{\mu\nu} + \frac{m^2}{2}(A_{\mu} + 
\partial_{\mu}\theta)(A^{\mu} + \partial^{\mu}\theta)\right )}.
\end{array}
\eqno(3.17)
$$
We have ignored a total derivative term in the second line in (3.17).
We thus arrive at the St{\"{u}}ckelberg Lagrangian$^{5}$. The $ 
\theta $ field is identified with the so-called St{\"{u}}ckelberg 
scalar, whose gauge transformation cancels that of the $ A_{\mu} $ 
field, thus making $ L $ invariant. Note that this $ L $ looks like 
the Proca Lagrangian, but here in (3.17) the fields $ (A_{\mu} + 
\partial_{\mu}\theta) $ are all gauge invariant. It may also be 
noted that the St{\"{u}}ckelberg Lagrangian in (3.17) goes back 
to the Proca Lagrangian (2.1) under the (unitary) gauge condition 
$ \theta = 0. $

A remark is in order at this stage. Using the Batalin-Fradkin method
a similar result has been obtained by Banerjee et al, Sawayanagi$^{6}$
and Kim et al$^{7}$. There the phase space is enlarged by introducing 
an extra canonical conjugate pair of fields. The extra field is 
identified with the St{\"{u}}ckelberg scalar, and additional terms 
in this extra field appear in the Hamiltonian to make it gauge 
invariant.

In contrast, we have found the St{\"{u}}ckelberg scalar {\em within} 
the original phase space itself. This is just the $ \psi^{\,\prime} 
(= \pi_0) $ 
of (3.8). As we have shown, gauge unfixing {\em does not} allow this
$ \psi^{\,\prime} $ to be put to zero. As a result extra terms in $ 
\psi^{\,\prime} $ appear in the gauge invariant Hamiltonian. These 
extra terms correspond to the additional terms appearing in the
BF gauge invariant Hamiltonian$^{6,7}$. Thus the $ \psi^{\,\prime}
$ (with rescaling) of the gauge unfixing method is just the extra 
field introduced in the BF method$^{6,7}$. Indeed this identification 
is confirmed when we go back to the second class Proca model. In the 
gauge unfixing method this is achieved by gauge fixing with $ 
\psi^{\,\prime} \approx 0 $, whereas in the BF method the extra field 
is put to zero.

We also mention that the St{\"{u}}ckelberg Lagrangian is manifestly 
Lorentz invariant, thus confirming the Poincar{\'{e}} algebra (3.13) 
that we obtained using modified Poincar{\'{e}} group generators.

\vspace{4mm}

The gauge theory of case (ii) can be related to another model too.
To see this, we rewrite the Hamiltonian $ {{\widetilde H}_c}^{{}^{^{\,
\prime}}} $ of (3.10) as
$$
{{\widetilde H}_c}^{{}^{^{\,\prime}}} = {\displaystyle \int d^3x~\left 
(\frac{\displaystyle {\vec \pi}^2}{\displaystyle 2} + 
\frac{\displaystyle F_{ij}F_{ij}}{\displaystyle 4} + F_0\,{\vec 
\nabla}\cdot{\vec \pi} - \frac{m^2}{2}F_0^2 + \frac{m^2}{2}F_i^2 \right 
)},
\eqno(3.18)
$$
where
$$
\begin{array}{rcl}
F_i & = & A_i - \frac{\displaystyle {\partial_i\pi_0}}{\displaystyle
m^2},\\
F_0 & = & A_0.
\end{array}
\eqno(3.19)
$$
Using (3.11), we see that $ F_0 $ and $ F_i $ are gauge invariant fields.
Thus the Hamiltonian $ {{\widetilde H}_c}^{{}^{^{\,\,\prime}}} $ in 
(3.10) involves gauge non-invariant fields, whereas the $ {{\widetilde 
H}_c}^{{}^{^{\,\,\prime}}} $ in (3.18) has only gauge invariant fields. 
In contrast to the $ A_0 $ and $ A_i $ having zero Poisson brackets
among themselves, we have here
$$
\begin{array}{rcl}
\{ F_0, F_0 \} & = & \{ F_i, F_j \} = 0,\\
\{ F_0(x), F_i(y) \} & = & \frac{\displaystyle 1}{\displaystyle
m^2}\,\partial_{ix}\delta(x-y).
\end{array}
\eqno(3.20)
$$
Thus the price one pays for considering gauge invariant fields is the 
non-zero PB in (3.20). Note that the above PBs among $ F_0, F_i $ are 
just the Dirac brackets (2.7) among the $ A_0, A_i $ fields in the 
original Proca system. We next define $ G_{\mu\nu} = 
\partial_{\mu}F_{\nu} - \partial_{\nu}F_{\mu}, $ and find that
$$
\partial_{\mu}G^{\mu\nu} = -m^2F^{\nu} + g^{0\nu}\,{\vec \nabla}^2
\chi^{\,\prime}.
\eqno(3.21)
$$
Thus modulo a term in $ \chi^{\,\prime}, $ eqn.(3.21) is similar to the
corresponding equation in the Proca model, $ \partial_{\mu}F^{\mu\nu}
= -m^2A^{\nu} $ which however involves gauge non-invariant fields.

Since (3.21) leads to $ \partial_{\mu}F^{\mu} = 0 $ the $ F_{\mu} $ 
fields can be written in terms of a gauge invariant antisymmetric 
tensor field $ A^{\mu\nu} $
$$
\begin{array}{rcl}
F_{\mu} & = & \frac{1}{2}\epsilon_{\mu\nu\lambda\sigma}\partial^{\nu}
A^{\lambda\sigma}\\
& = & \frac{1}{6}\epsilon_{\mu\nu\lambda\sigma}G^{\nu\lambda\sigma},
\end{array}
\eqno(3.22)
$$
where we have the totally antisymmetric quantity $ G^{\nu\lambda\sigma}
= \partial^{[\nu}B^{\lambda\sigma]}. $ The Hamiltonian $ 
{{\widetilde H}_c}^{{}^{^{\,\prime}}} $ now becomes
$$
\begin{array}{rcl}
{{\widetilde H}_c}^{{}^{^{\,\prime}}} & = & {\displaystyle \int d^3x~
\left (\frac{\displaystyle \pi_i^2}{\displaystyle 2} + 
\frac{\displaystyle F_{ij}F_{ij}}{\displaystyle 4} + 
F_0(\partial_i\pi_i) - \frac{\displaystyle m^2}{\displaystyle 
8}(\epsilon_{ijk}\partial^iA^{jk})^2 + 
\frac{\displaystyle m^2}{\displaystyle 
8}(\epsilon_{i\nu\lambda\sigma}\partial^{\nu}A^{\lambda\sigma})^2 
\right )}\\
& = & \rule{0mm}{7mm} {\displaystyle \int d^3x~\left 
(\frac{\displaystyle \pi_i^2}{\displaystyle 2} + \frac{\displaystyle 
F_{ij}F_{ij}}{\displaystyle
4} + F_0(\partial_i\pi_i) + \frac{\displaystyle m^2}{\displaystyle 12}
G_{ijk}G^{ijk} + \frac{\displaystyle m^2}{\displaystyle 4}G_{0jk}G^{0jk}
\right )}
\end{array}
\eqno(3.23)
$$
with $ \epsilon_{ijk} = \epsilon_{0ijk}. $Note that (3.23) involves 
only gauge invariant fields.

It is more interesting to consider gauge non-invariant antisymmetric
tensor fields. Recall that the gauge invariant Hamiltonian $ 
{{\widetilde H}_c}^{{}^{^{\,\prime}}} $ was first obtained as a series 
(3.10) in the $ \pi_0, $ which was later redefined to be the 
St{\"{u}}ckelberg scalar $ \theta (= \frac{\!-\displaystyle 
\pi_0}{\displaystyle m^2}). $ Instead of a scalar field,
we can introduce a tensor field, while still retaining gauge 
invariance. For this, we use (3.19) and (3.22) to write
$$
\begin{array}{rcl}
\partial_i\,\pi_0 & = & m^2(A_i - \frac{1}{2}\epsilon_{i\mu\nu\lambda}
\partial^{\mu}A^{\nu\lambda})\\
& = & \frac{1}{2}\epsilon_{ijk}\pi^{jk},\\
\pi_{jk} & = & m^2(\epsilon_{jkm}A^m + G_{0jk})
\end{array}
\eqno(3.24)
$$
The Hamiltonian of (3.10) now becomes
$$
{{\widetilde H}_c}^{{}^{^{\,\,\prime}}} = H_c + {\displaystyle \int d^3x~
\left (\frac{\displaystyle 1}{\displaystyle 4m^2}\pi_{ij}\pi^{ij} +
\frac{1}{2}\epsilon_{ijk}A^i\pi^{jk}
\right )}
\eqno(3.25)
$$
where $ H_c $ is the Proca Hamiltonian (2.2). Thus in place of a 
(finite) series in a scalar field, we now have $ {{\widetilde 
H}_c}^{{}^{^{\,\,\prime}}} $ to be a series in the tensor field $ \pi_{ij}. $
The gauge theory involving $ \theta $ (or $ \pi_0 $) had the $ A_{\mu} $
field interacting with the $ \theta $ field; here $ A_{\mu} $ 
interacts with an antisymmetric tensor field. Note that the 
unitary gauge $ \pi_{ij} = 0 $ takes $ {{\widetilde 
H}_c}^{{}^{^{\,\,\prime}}} $ back to the Proca Hamiltonian $ H_c; $ 
this, from (3.24) is just the $ \pi_0 = 0 $ used earlier.

The Hamiltonian $ {{\widetilde H}_c}^{{}^{^{\,\,\prime}}} $ in (3.25) 
is invariant under gauge transformations generated by $ \left (
\frac{\displaystyle -{{\vec \nabla}\cdot{\vec \pi}}}{\displaystyle 
m^2} + \frac{1}{2}\epsilon_{ijk}\partial^{\,i}A^{jk} \right ), $ 
which is obtained from $ \chi^{\,\prime} = \frac{1}{m^2}(-\partial_i\pi_i 
+ m^2A_0) $ using (3.22). The fields $ A^{\mu\nu} $ are gauge 
invariant, from (3.22). The tensor $ \pi_{jk} $ however is not; using
$$
\{ A^{ij}(x), \pi_{mn}(y) \} = (\delta^i_{\,m}\delta^j_{\,n}
-\delta^i_{\,n}\delta^j_{\,m})\;\delta(x-y),
\eqno(3.26)
$$
we find the variation $ \pi^{jk} \rightarrow \pi^{jk} - 
\epsilon^{ijk}\partial_i\mu, $
where $ \mu $ is the transformation parameter. Note that the relation
(3.26) and the above variation of $ \pi_{jk} $ are consistent
with (3.24) and the variation (3.11) of $ \pi_0. $

The Hamiltonian (3.25) is very similar to the one obtained by 
Sawayanagi$^{6}$, who has used the BF method. The extra fields
introduced were just the tensor field $ \pi_{ij} $ and $ (\frac{1}{2}
\epsilon_{ijk}\partial^iA^{jk} - A_0), $ which were used to write 
down the gauge invariant Hamiltonian as a (finite) series (this 
Hamiltonian$^{6}$ has an extra term involving 
$(\,\frac{1}{2}\epsilon_{ijk}\partial^iA^{jk} - A_0), $ which is 
zero in our case, see (3.22)). Our result (3.25) however is 
obtained {\em within} the original phase space.

The Hamiltonian $ {{\widetilde H}_c}^{{}^{^{\,\,\prime}}} $ may not 
lead to a manifestly Lorentz invariant Lagrangian involving the Maxwell
and tensor fields (we have not considered Lorentz invariance in phase
space here, since it has already been verified in (3.13)). We can 
however write down such a Lagrangian$^{6,8}$ which gives the
Hamiltonian (3.25),
$$
{{\widetilde L}}^{{}^{^{\,\,\prime}}} = {\displaystyle \int d^3x~\left 
(~-\frac{1}{4}F_{\mu\nu}^2 - \frac{\displaystyle m^2}{\displaystyle 6}
\epsilon_{\mu\nu\rho\sigma}A^{\mu}G^{\nu\rho\sigma} + 
\frac{\displaystyle m^2}{\displaystyle 12}G_{\mu\nu\rho}G^{\mu\nu\rho} 
\right )},
\eqno(3.27)
$$
with $ G_{\mu\nu\alpha} $ antisymmetrised in all the indices. The phase 
space involves $ A^{\mu}, A^{\mu\nu} $ and the canonical momenta
$$
\begin{array}{rcl}
\pi_i & = & -F_{0i}   ~~~~~~~~~~~~~~~~~~~~~~~~~\pi_{ij} = 
m^2(\epsilon_{ijk}A^k + G_{0ij})\\
\pi_0 & = & 0       ~~~~~~~~~~~~~~~~~~~~~~~~~~~~~~\pi_{0i} = 0
\end{array}
\eqno(3.28)
$$
with the second line giving the primary constraints. The canonical
Hamiltonian is 
$$
\begin{array}{rcl}
H_{\em inv} & = & {\displaystyle \int d^3x~ \left (\frac{1}{2}\pi_i^2 +
\frac{\displaystyle m^2}{\displaystyle 2}A_i^2 + \frac{1}{4}F_{ij}^2
+ \frac{\displaystyle 1}{\displaystyle 4m^2}\pi_{ij}^2 + \frac{1}{2}
\epsilon_{ijk}A_i\pi_{jk} \right. }\\
&   & ~~~~~~~~~~~~~~~~{ \left. + \frac{\displaystyle m^2}{\displaystyle 
12}G_{ijk}^2 - A_0(-{\partial_i\pi_i} + \frac{\displaystyle 
m^2}{\displaystyle 2}\epsilon_{ijk}\partial^iA^{jk}) + 
A_{0j}\partial_i\pi_{ij} \right )}.
\end{array}
\eqno(3.29)
$$
The time independence of the primary constraints in (3.28) yield the
secondary constraints,
$$
\begin{array}{rcl}
-{\vec \nabla}\cdot{\vec \pi} + \frac{\displaystyle m^2}{\displaystyle
2}\epsilon_{ijk}\partial^iA^{jk} & = & 0,\\
\partial_i\pi_{ij} & = & 0.
\end{array}
\eqno(3.30)
$$
Modulo these constraints and using (3.22), we find that (3.29) is just
the Hamiltonian $ {{\widetilde H}_c}^{{}^{^{\,\,\prime}}} $ of (3.25). 
Note that the constraints in (3.28) and (3.30) are all first class, 
showing that (3.27) describes a gauge theory.

\vspace{5mm}

{\noindent {\large{\bf{4. {\underline {Conclusion}}}}}}

\vspace{3mm}

In this paper we have revealed gauge symmetries inherently present 
in the gauge non-invariant Proca model. We have used the Gauge 
Unfixing method, the central object of which is the projection 
operator. We have shown that this operator defines the gauge theory 
by projecting all relevant quantities (constructed initially on the 
second class constrained surface) onto a first class constrained 
surface. This projection operator is not unique; there are two 
different operators, which implies two different gauge theories. 
We have shown that one of these results in a trivial gauge invariance, 
and the other gives a non-trivial one. In each of these gauge theories 
we have verified Poincar{\'{e}} invariance by (necessarily) modifying 
the Poincar{\'{e}} generators of the original Proca model.

{}For the first gauge theory (case[i]), the corresponding Lagrangian 
is not manifestly Lorentz invariant (even though in phase space Lorentz 
invariance is confirmed). As for the second gauge theory the passage 
to the Lagrangian formulation results in a manifestly Lorentz invariant 
Lagrangian, thus confirming the Lorentz invariance shown in phase 
space. Further this Lagrangian is just the St{\"{u}}ckelberg 
Lagrangian, which was proposed quite sometime back by 
St{\"{u}}ckelberg$^{5}$ by adding extra terms in an extra 
(St{\"{u}}ckelberg) field directly to the Proca Lagrangian. From 
the constraints point of view our method is thus consistent with 
the St{\"{u}}ckelberg formulation.

The St{\"{u}}ckelberg Lagrangian has also been derived using 
the Batalin-Fradkin
(BF) method$^{6,7}$, which is formulated by enlarging the phase
space. We emphasize that the Gauge Unfixing method derives this 
Lagrangian {\em without} any extension of the phase space (similar
conclusions have been arrived at for other systems also --- the 
abelian Chern-Simons theory and the abelian chiral Schwinger 
model$^{3}$). Thus we have a connection between the two methods.

We have also shown that the gauge theory of case(ii) leads to 
another formulation, that of the Maxwell field interacting with
an antisymmetric tensor field. Whereas this was shown by Banerjee
and Sawayanagi$^{6}$ to arise
in an extended phase space, our analysis here shows that the original
phase space is sufficient to reproduce such a theory.

It would be interesting to see how well the method works for the 
non-abelian Proca model. In this model, it is not just the $ 
\pi_0 \approx 0 $ constraints and the Gauss law constraints which are 
second class with each other, but the Gauss law constraints are second 
class among themselves. It may be possible to use the Gauge Unfixing
method (under certain conditions, see [3]) for these systems too.
Work is in progress in this direction.

\vspace{6mm}
{\noindent\large{\bf{Acknowledgements}}}

\vspace{3mm}

We wish to thank the Council for Scientific and Industrial Research,
New Delhi for financial assistance for this work. We also thank Dr B A
Kagali for constant encouragement, and Prof M N Anandaram (Bangalore 
University) and Centre for Theoretical Studies (IISc, Bangalore) for 
providing computer facilities. We also thank the referee for 
insightful comments and suggestions.

\vspace{4mm}
{\noindent{\bf{Note Added}}}~ After completion of this work, we became
aware of another paper$^{9}$, where the conversion of the abelian Proca
model into a first class system has been discussed.
\vspace{7mm}

{\noindent \large {\bf{ APPENDIX}}}

\vspace{3mm}

The projection operator $ I\!\!P = :e^{-{\displaystyle \int d^3x 
~\psi{\hat \chi}}}: $ has the following properties:
$$
I\!\!{P^2} \cong I\!\!P
\eqno(A.1)
$$
$$
\!\!\!I\!\!P(bB + c\,C) = b{\widetilde B} + c{\,\widetilde C} 
\eqno(A.2)
$$
$$
\!\!{\hat \chi}I\!\!P \cong 0
\eqno(A.3)
$$
$$
I\!\!P(\psi) = {\widetilde \psi} \cong 0
\eqno(A.4)
$$
$$
\!\!({\widetilde {BC}}) = {\widetilde B}{\widetilde C}
\eqno(A.5)
$$
$$
{}~~~\{{\widetilde B}, {\widetilde C}\} \cong I\!\!P\left (\{ B, 
C\}_{DB}\right )
\eqno(A.6)
$$
$$
\{{\widetilde B},\{{\widetilde C}, {\widetilde D}\}\} + 
\{{\widetilde C},\{{\widetilde D}, {\widetilde B}\}\} + \{{\widetilde 
D},\{{\widetilde B},{\widetilde C}\}\} \cong 0
\eqno(A.7)
$$
where the symbol $ \cong $ implies equality on the surface defined by 
only $ \chi \cong 0. $ The proofs for the above properties can be found 
in [3].

\vspace{7mm}

{\noindent {\large{\bf{References}}}}

\vspace{5mm}
\begin{tabular}{lp{6.4in}}
{{\bf (1)}} & P A M Dirac, {\em Lectures on Quantum Mechanics, Belfer
	      Graduate School of}\\
	    & {\em Science, Yeshiva University, New York, 1964};\\
	    & A J Hanson, T Regge and C Teitelboim, {\em Accademia 
	      Nazionale dee Lincei,}\\
	    & {\em Rome, 1976}.\\
{{\bf (2)}}  & I A Batalin and E S Fradkin, {\em Phys. Lett. \bf{B 180}
	      (1986) 157; ~ Nucl. Phys.}\\
	    & {\em \bf{B 279} (1987) 514}; \\
	    & I A Batalin and I V Tyutin, {\em Int. J. Mod. Phys. \bf{
	      A 6} (1991), 3255 and references therein}.\\
{{\bf (3)}} & P Mitra and R Rajaraman {\em Ann. Phys. (N Y) \bf{203}
	      (1990) 137, 157}; \\
	    & R Anishetty and A S Vytheeswaran {\em J. Phys. \bf{A 26} 
	      (1993) 5613};\\
	    & A S Vytheeswaran {\em Ann. Phys. (N Y) \bf{236} (1994) 
	      297}.\\
{{\bf (4)}} & L D Faddeev {\em Theor. Math. Phys. \bf{1} (1969) 1}; \\
	    & E S Fradkin and G A Vilkovisky {\em Phys. Lett. \bf{B 55} 
	      (1975) 224}; \\
	    & M Henneaux {\em Phys. Rep. \bf{126} (1981) 1}.\\
{{\bf (5)}} & E C G St{\"{u}}ckelberg {\em Helv. Phys. Act. \bf{30} 
	      (1957) 209.}\\
{{\bf (6)}} & N Banerjee and R Banerjee {\em Mod Phys Lett. \bf{A11}
	      (1996) 1919}; \\
	    & H Sawayanagi {\em Mod. Phys. Lett. \bf{A 10} (1995) 813.}\\
{{\bf (7)}} & Y W Kim, M I Park, Y J Park, S J Yoon {\em hep-th/9512110; 
	      To appear in}\\
	    & {\em International Journal of Modern Physics A}. \\
{{\bf (8)}} & T J Allen, M J Bowick and A Lahiri {\em Mod. Phys. Lett. 
	      \bf{A 6} (1991) 559;}\\
	    & A Lahiri {\em Mod. Phys. Lett. \bf{A 8} (1993) 2403.}\\
{{\bf (9)}} & Omar F Dayi, {\em Phys. Lett. \bf{B 210} (1988) 147,}\\
            & C Bizdadea and S O Saliu {\em EuroPhys. Lett. \bf{32}
              (1995) 307-312.}
\end{tabular}

\end{document}